\documentclass[amsmath,amssymb,aps,twocolumn]{revtex4-2}
\usepackage{graphicx}
\usepackage{dcolumn}
\usepackage{bm}
\usepackage{xcolor}
\begin{document}
\title{Anomalous velocity distributions in slow quantum-tunneling chemical reactions}
\author{Christian Beck$\,^{1,2,*}$}
\author{Constantino Tsallis$\,^{3,4,5}$}
\affiliation{$\!^1$Centre for Complex Systems, Queen Mary University of London, School of Mathematical Sciences, Mile End Road, London E1 4NS, UK}
\affiliation{$\!^2$International Excellence Fellow at Karlsruhe Institute of Technology, D-76131 Karlsruhe, Germany}
\affiliation{$\!^3$Centro Brasileiro de Pesquisas F\'{\i}sicas and National Institute of Science and Technology of Complex Systems, Rua Xavier Sigaud 150, 22290-180, Rio de Janeiro RJ, Brazil}
\affiliation{$\,^4$Santa Fe Institute, 1399 Hyde Park Road, Santa Fe, 87501 NM, USA}
\affiliation{$\,^5$Complexity Science Hub Vienna,  Josefst\"adter Strasse 39, 1080 Vienna, Austria}

\email{c.beck@qmul.ac.uk}

\begin{abstract}
Recent work [Wild et al., Nature 615, 425 (2023)] has provided an experimental  break-through in the realization of a quantum-tunneling reaction involving a proton transfer. 
The reaction $D^-+H_2 \to H^-+HD$ has an extremely slow reaction rate as it can happen only via quantum tunneling, thus requiring an extremely large density of the reactants in the ion trap. At these high densities strong deviations from Maxwell-Boltzmann statistics are observed.
Here we develop a consistent generalized statistical mechanics theory for the above nonequilibrium situation involving quantum effects at high densities. The trapped ions are treated in a superstatistical way and a $q$-Maxwellian velocity distribution with a universal dependence of the entropic index $q$ on the density $n$ of the buffer gas is derived. 
We show that the velocity distribution of the ions is non-Maxwellian, more precisely $q$-Gaussian, i.e., $p(v) \propto v^2 [1+(q-1)\tilde{\beta} v^2]^{1/(1-q)}$,  with entropic index $q>1$ depending on the density $n$ of $H_2$ molecules, in excellent agreement with the experimental observations of Wild et al. 
Our theory
also makes predictions on the statistics of temperature fluctuations in the ion trap which can be tested in future experiments. 
Through the superstatistical approach,  we obtain an analytical expression for $q(n)$ which is consistent with the available experimental data, and which yields $\lim_{n\to 0}q(n)=1$, i.e. recovering the Maxwell-Boltzmann distribution in the ideal gas limit, as well as $\lim_{n\to\infty}q(n)=7/5$. 
\end{abstract}

\maketitle

Ordinary chemical reactions are usually described by ordinary statistical mechanics, that is to say the reactants obey Boltzmann-Gibbs (BG) statistics.
The time evolution of ordinary chemical reactions is described by the Arrhenius law, which is in turn grounded on BG statistical mechanics. Consequently, when the reaction occurs, say in a gas in a container of large volume, the distributions of velocities of the various chemical components are of the celebrated Maxwellian (Gaussian) form. In a typical picture for such a reaction, the system is initially at the bottom of a local minimum of a potential energy well and, by overcoming an activation barrier, it ends up in a global minimum. 
The situation, however, is quite different in modern state-of-the art experiments based on ion traps. Here the assumption of Maxwell-Boltzmann statistics has been shown to be inherently violated \cite{Voe2009,Asvani2009,Rouse2017,Noetzold2020}. This is due to the fact that the reactants are usually confined in a very small volume, so that temperature fluctuations occur and nonequilibrium effects play an important role.
Also, the usual assumption of a small density (which makes kinetic theory applicable) can be inherently violated, so that intrinsic quantum effects play an important role. Moreover, the ions are constantly driven by an external field, so that their temperature, due to rf (radio frequency) driving, is usually larger than that of the embedding neutral buffer gas, a nonequilibrium situation. There is also heat loss in the system due to the possibility of some ions vanishing from the trap.

As an example of a recent ground-breaking experiment in this direction, where the assumption of Maxwell-Boltzmann statistics is inherently violated, let us consider the experiment described in \cite{Wildnature} (for other, more general experiments, see e.g. the recent review \cite{deiss-review}).
This experiment deals with the reaction $D^- + H_ 2 \to H^- + HD$ and it is ground-breaking because it reaches a much higher density of reactants than previous ion-trap experiments. In the above reaction a proton transfer occurs, however the rate constant is extremely small so that a very high density of reactants is needed to see a measurable effect. This high density was achieved in \cite{Wildnature} and the reaction was actually experimentally observed for the first time. {Consequently, due to the high density, very sensitive quantum effects are dominant in this ultra confined driven system at low temperatures. Moreover, quantum tunneling transitions rather than classical transitions can be experimentally studied for this system, which provide the dominant pathway to enable the reaction at smaller densities \cite{McMahon2003,ShannonBlitzGoddardHeard2013,Tiznitietal2014,Yang2019,Yuen2018}.} 

Interestingly enough, but understandable due to the small volume and high density of the driven system, the distribution of velocities $v$ of the ions was shown (see e.g.\ Fig.~3c of  \cite{Wildnature}) to be anomalous, in the sense that it has a $q$-Maxwellian form, instead of the {\em a priori} expected Maxwellian form $p(v) \propto v^2 e^{-\frac{\beta}{2}m  v^2}$. The three-dimensional $q$-Maxwellian velocity distribution is given by 
\vspace{-0.3cm}
\begin{equation}
p_q(v) \propto v^2 e_q^{-\frac{\beta_q}{2}m  v^2} \,,
\label{qGaussian}
\end{equation}
where the $q$-exponential function is defined as
\begin{equation}
e_q^{\,z} \equiv [1+(1-q)z]^{\frac{1}{1-q}} 
= \frac{1}{[1-(q-1)z]^{\frac{1}{q-1}}} \;\;\;(q \ge 1; \, e_1^{\, z}=e^z)\,.
\label{qexponential}
\end{equation}
These distributions maximize the $q$-entropies $S_q$ \cite{Tsallis1988} subject to suitable constraints. In the following we will derive the above distributions in the given experimental context. A particular highlight will be
that we will obtain a theoretical prediction how the entropic index $q$ 
in the generalized statistical mechanics description 
\cite{Tsallis1988,Tsallis2023book, beck-review} 
depends on the
density $n$ of $H_2$ molecules. Our approach will be based on superstatistics \cite{Beck2001, BeckCohen2003, BCS} and ultimately
we will derive an area-law for the entropic index $q-1$.
The relevance of superstatistics in the context of ion trap experiments was previously emphasized in \cite{Rouse2017}. Here we further develop the theory and extend it to the case of large densities, which is necessary to understand experiments such as the recent one described in \cite{Wildnature}.
Our main result is a generally applicable thermodynamic superstatistical theory that correctly describes the behaviour of quantum tunneling reactions in ion trap experiments 
at very high and very small densities alike.

\begin{figure}
\centering
    \includegraphics[width=6cm]{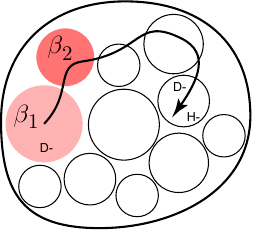}
    \caption{The basic idea of superstatistics. The spatial region in which the reactants are contained consists of several subregions of inverse temperature $\beta_1, \beta_2, \beta_3, \ldots$ through which the molecules move. Each subregion is temporarily in local thermal equilibrium at local temperature $\beta_i^{-1}$. {For a superstatistical description to be valid, there must be a clear time scale separation in the system, i.e.\ the relaxation to local equilibrium must be faster than the typical time scale of temperature variations. See \cite{straten} for details on how to extract these parameters from experimentally measured time series.}}
    \label{betacells}
\end{figure}

Our aim in the following is to obtain a nonequilibrium statistical physics understanding how the entropic index $q$ depends on the density $n$ of the $H_2$ particles involved. Clearly, as the quantum-tunneling reaction has a very small reaction rate, a very high density $n$ is required to produce a measurable effect. Keeping the number of particles constant, a very high density corresponds to a very small volume $V$. It is here where the superstatistical approach comes into play \cite{Beck2001,BeckCohen2003,BCS}:
In a very small volume we expect temperature not to be constant but there will be temperature fluctuations, meaning $\beta=1/kT$ becomes a random variable (see Fig.~\ref{betacells}).

We may associate these temperature fluctuations either with classical fluctuations (heat losses through the surface) or with quantum fluctuations (uncertainties of the temperature) that are allowed due to the uncertainty relation. In the $\chi^2$ superstatistics approach, introduced in \cite{BeckCohen2003} and applied to ion traps in \cite{Rouse2017}, one assumes the following form for the inverse temperature random variable:
\begin{equation}
    \beta = \sum_{i=1}^\mu X_i^2 \label{n-degrees}
\end{equation}
Here the $X_i$ are Gaussian random variables with mean 0, which are squared in order to produce a positive contribution to $\beta=1/kT$. It must be a positive contribution since $\beta$ must always be positive. The number $\mu$ is an effective number of degrees of freedom that contribute to the temperature fluctuations. 

Let us now consider a very small volume (equivalent to a very large density), where the density is so large that quantum effects start to dominate and nonequilibrium effects are dominant, meaning there are big temperature fluctuations. We expect the fluctuations of $\beta$ to be dominated by the heat loss flow through the surface of the given small volume, for the three spatial dimensions in which the experiment is performed. In addition, we may allow for an additional explicit time dependence of the temperature in a given volume.
We thus arrive at the result that at the highest experimentally possible densities $n_{max}$ that are achievable in the experimental setup the number $\mu$ of degrees of freedom  describing the temperature fluctuations is given by $\mu=4$,
with 3 degrees of freedom corresponding to space and 1 degree to time, so that
\begin{equation}
\beta= X_1^2+X_2^2 +X_3^2 +X_4^2.
\end{equation}
This formula is only valid at the highest experimentally achievable densities $n_{max}$, i.e. the smallest possible volume. If the density decreases,
which is equivalent to the volume increasing (keeping the particle number constant), we then expect more degrees of freedom to enter the system, as heat can flow in all kinds of local directions, but only the losses through the surface that includes all reactants are relevant. Because of this surface effect, we expect the
number $\mu$ of degrees of freedom to grow as
\begin{equation}
    \mu \sim V^\frac{2}{3} \sim n^{-\frac{2}{3}}
\end{equation}
since the surface of the system (through which the heat losses occur) generally goes like volume $V$ to the power $2/3$.
At the highest possible density $n_{max}$ achievable in the experiment we
have $\mu_{min}=4$. Thus, as $n \sim V^{-1}$, we arrive at the scaling formula
\begin{equation}
    \mu =3 \cdot \left( \frac{V}{V_{min}} \right)^\frac{2}{3} +1 =3 \cdot \left( \frac{n_{max}}{n} \right)^\frac{2}{3}+1 \,, \label{mu}
\end{equation}
which is a formula that provides us with the effective
degrees of freedom $\mu$ of the temperature fluctuations
as a function of the density $n$ of the quantum reactants.

Let us now do superstatistics \cite{Beck2001,BeckCohen2003,BCS} (i.e.\ we consider averaging over the fluctuating inverse temperatures, weighted with a probability density function $f(\beta)$ to observe a particular $\beta$,
see End Matter section for more details).
Assume we know the inverse temperature $\beta$ as given in a particular spatial region at a particular time. Then at this point we have
local Maxwell Boltzmann statistics
\begin{equation}
    p(E|\beta)=C E e^{-\beta E} \label{t1}
\end{equation}
where $E=\frac{1}{2}mv^2$ is the kinetic energy of the reactant, $p(E|\beta)$ is the conditional probability to observe an energy state $E$ given the local environment with inverse temperature $\beta$. The normalization constant $C$ is readily obtained from
\begin{equation}
    \int_0^\infty p(E| \beta) dE=1 
\end{equation}
to be 
\begin{equation}
    C= \beta^2. \label{t2}
\end{equation}
Due to the temperature fluctuations in the system, we need to average over all states of the inverse temperature variable $\beta$,
weighted with the probability density $f(\beta)$ to observe a particular $\beta$, to obtain the marginal distribution
\begin{equation}
    p_q(E)= \int_0^\infty d\beta \; f(\beta) p(E|\beta). \label{q-exp}
\end{equation}
In our case, from the assumption that $\beta$ is given by
eq.~(\ref{n-degrees}), we immediately identify $f(\beta)$ as the $\chi^2$ distribution with $\mu$ degrees of freedom:
\begin{equation}
    f(\beta) = \frac{1}{\Gamma(\frac{\mu}{2})} \left( \frac{\mu}{2 \beta_0} \right)^{\frac{\mu}{2}} \beta^{\frac{\mu}{2}-1} \exp
    \left( - \frac{\mu}{2} \frac{\beta}{\beta_0} \right) \label{t3}
\end{equation}
where
\begin{equation}
    \beta_0= \langle \beta \rangle = \int_0^\infty \beta f(\beta) d\beta =\mu \langle X_i^2 \rangle
\end{equation}
is the average inverse temperature in the system, and the variance
of inverse temperature fluctuations is given by
\begin{equation}
    \langle \beta^2 \rangle - \beta_0^2 = \int_0^\infty \beta^2 f(\beta) d \beta - \beta_0^2=\frac{2}{\mu} \beta_0^2. \label{20}
\end{equation}
The distribution (\ref{t3}) represents one of 3 possible universality classes that are usually considered in the superstatistics approach \cite{BCS}, and its relevance in the context of ion trap experiments was recently derived in \cite{Rouse2017} where it was shown that it arises from a recurrence relation for temperature due to subsequent collisions of the ions.

Let us now explicitly do the integration in eq.~(\ref{q-exp}),
using eqs.~(\ref{t1}), (\ref{t2}), and (\ref{t3}).
The result of a short calculation (see End Matter section for details) is
\begin{equation}
    p_q(E)=\int_0^\infty d\beta f(\beta) \beta^2 Ee^{-\beta E} \sim E e_q^{- \tilde{\beta} E},
\end{equation}
where
\begin{equation}
    q=1+\frac{2}{\mu +4} \label{q}
\end{equation}
and
\begin{equation}
    \tilde{\beta}= \frac{\beta_0}{1-2(q-1)}. \label{bq}
\end{equation}
In other words, the integration over the fluctuating temperature states of the system containing the reactants quite naturally leads to $q$-exponential distributions. We have thus clarified how the nonequilibrium situation of the small volume of reactants with a high density and temperature fluctuations (as sketched in Fig.~\ref{betacells}) leads to the actually observed $q$-Maxwellian distributions.

\begin{figure}[ht]
\centering
\includegraphics[width=6cm]
{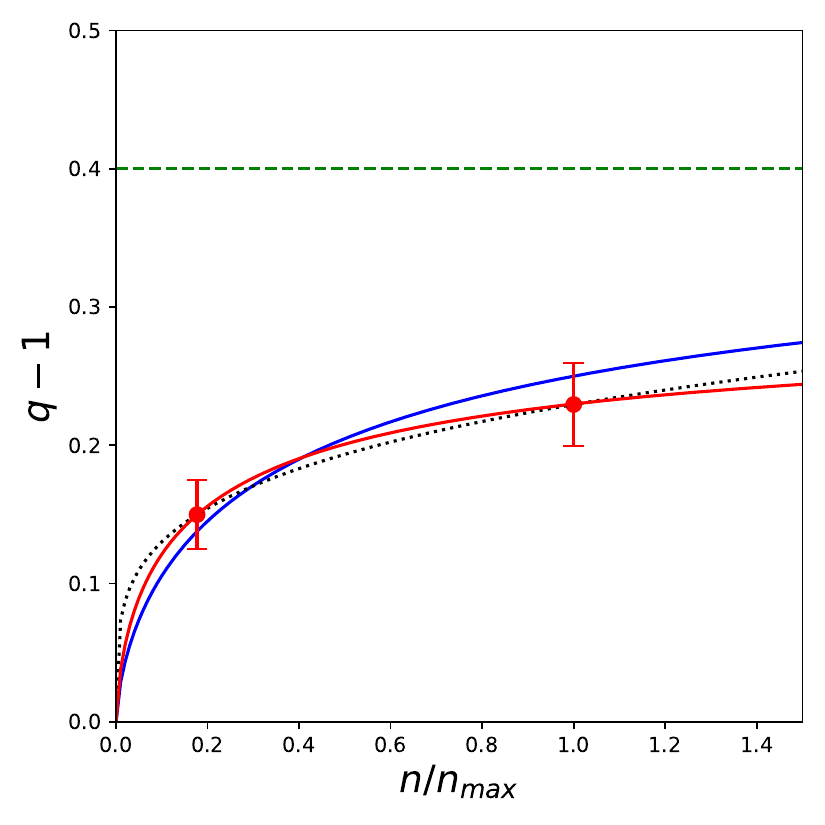}
\caption{Entropic index $q-1$ describing quantitatively the deviation from BG statistics as a function of the density $n$ of reactants. The two red data points correspond to the experimental values as provided in Fig. 3(c) of \cite{Wildnature}. 
The dotted curve corresponds to the heuristically conjectured $(q-1) \propto n^{1/4}$ $(n\to 0)$ of \cite{Tsallis2023}; this curve exactly passes through the two (red) experimental points by construction.  The blue curve corresponds to eq. (\ref{qn}) $[(q-1)\propto n^{2/3}$ $(n\to 0)]$ analytically deduced within the present superstatistical approach,
and expected to hold under idealized conditions. For the blue curve, $\lim_{n\to\infty} q(n)=7/5$ (green dashed line); for $q \ge 7/5$, $\langle v^2 \rangle$ diverges. 
The red curve is the theoretical prediction of eq.~(\ref{q-general-a}) and (\ref{xxx}), adjusting the two non-universal fitting constants $C_1=2.14$ and $C_0=2.56$ to their optimum values for the given experiment.}
\label{figureall}
\end{figure}

The dependence of the $q$-parameter on the density $n$ of reactants
can now be predicted by combining eq.~(\ref{q}) with eq.~(\ref{mu}). We obtain the formula
\begin{equation}
    q-1= \frac{2}{\mu(n)+4}= \frac{2}{3 \left( \frac{n_{max}}{n} \right)^\frac{2}{3} +5}.  \label{qn}
\end{equation}
This formula, within the estimated error bars, is in good agreement with the experimental results (see blue curve in Fig.~\ref{figureall}). Let us discuss a few special cases.

\begin{enumerate}
\item For $n \to 0$ we have $q \to 1$. This makes physical sense. For small densities of reactants, ordinary statistical mechanics described by $q=1$ is recovered.

\item For $n=n_{max}$, we obtain $q=\frac{5}{4} =1.25$. This is the case where the density $n$ takes on the largest achievable value $n_{max}$ that is reachable in the experiment. In this case there are strong temperature fluctuations in the system, and heat losses occur through the 3 directions of space of the smallest possible elementary reaction volume $V_{min}$. We also allow for an explicit time dependence of the temperature, making the total number of degrees of freedom contributing to the fluctuations of temperature to be $\mu=4$.

\item Although not experimentally achievable, we may formally consider the limit $ n \to \infty$ of infinite density. In this case formally the system is reduced to a 1-dimensional point system that is infinitely concentrated and has only one degree of freedom left, the explicit time dependence of the temperature.
This limit leads to $\mu =1$ and the resulting $q$-value is $q=q_\infty=\frac{7}{5}=1.4$ .

\end{enumerate} 

Interesting enough, $q=\frac{7}{5}$ as discussed under point 3. above, is precisely the limit value up to which the variance of the velocity $v$ of the particles still exists. This can be seen as follows: We have for the $q$-Maxwellian distributions
\begin{equation}
    p_q(v) \sim v^2 e_q^{-\tilde{\beta} \frac{1}{2}mv^2}\sim v^{2-\frac{2}{q-1}} \;\;(v \to \infty).
\end{equation}
Hence for the integral defining the second moment of velocity $v$ 
\begin{equation}
    M_2= \langle v^2 \rangle =\int_0^\infty v^2 p_q(v) dv
\end{equation}
the integrand decays as $v^{4-\frac{2}{q-1}}$ for large $v$, and hence
the 2nd moment $M_2$ only exists if
\begin{equation}
    q<q_\infty =\frac{7}{5}
\end{equation}
as for $q$-values smaller than $\frac{7}{5}$ the integrand decays stronger than $v^{-1}$ and is thus
integrable. We obtain the interesting result that our value of $q_\infty=\frac{7}{5}$, obtained previously from the $n \to \infty$ asymptotics of the formula (\ref{qn}), coincides precisely with the point up to which the variance of the velocity distribution exists. 

Next, we deal with the effective temperature parameter of the system of quantum reactants. This is described by the parameter $\beta_q$ in eq.~(\ref{qGaussian}), and in the superstatistics approach the same parameter is usually denoted as $\tilde{\beta}$. The parameter $\beta_q$ has a non-trivial $q$-dependence (and thus density dependence) which we can derive by combining
eq.~(\ref{bq}) and eq.~(\ref{q}) with eq.~(\ref{mu}).
We obtain by putting eq.~(\ref{mu}) into eq.~(\ref{bq}) 
\begin{equation}
    \beta_q\equiv \tilde{\beta}=\beta_0 \cdot \frac{3 \left( \frac{n_{max}}{n}\right)^\frac{2}{3}+5}{3 \left( \frac{n_{max}}{n} \right)^\frac{2}{3}+1}.
\end{equation}
This yields an interesting experimentally testable prediction for the generalized statistical mechanics
describing the slow quantum-tunneling reaction in the ion trap experiment: The effective inverse temperature parameter in the  $q$-distribution becomes density dependent.
Whereas for $n\to 0$ we have $\beta_q=\beta_0$, i.e. ordinary statistical mechanics, for $n \to n_{max}$ we obtain $\beta_q \to 2 \beta_0$. If we interpret $\beta_0^{-1}$ as the physical average temperature of the ions
(which are in fact in a non-equilibrium situation with fluctuating temperature) and if we then fix the temperature parameter $\tilde{\beta}=\beta_q$ to be used in the $q$-distribution from the temperature of the embedding $H_2$ gas (which is  having no temperature fluctuations) then the temperature $\beta_0^{-1}$ can be interpreted as the average temperature of the ions and $\beta_q^{-1}$ as the fixed temperature of the embedding $H_2$ gas. Clearly, the ion temperature is larger than the $H_2$ bath temperature for any $q>1$. This is a direct consequence
of eq.~(\ref{q}), which we may write in terms of the temperature parameters
as 
\begin{equation}
T_{ion}=\frac{1}{1-2(q-1)} T_{bath}.
\end{equation}
For example, for $n=n_{max}$ (equivalent to $q=\frac{5}{4}$) we obtain the result that the ion temperature is larger by a factor 2 than the $H_2$ bath temperature. This tendency is in agreement with observations: The ion temperature is always observed to be bigger than the bath temperature in the ion trap experiments. The physical interpretation is that this is due to rf heating.

We may finally use standard formulas from superstatistics to estimate the size
of temperature fluctuations for the ion gas. Eq.~(\ref{20}) implies that the degrees of freedom $\mu$ fix the variance of inverse temperature fluctuations according to the general formula
\begin{equation}
    \frac{ \langle \beta^2 \rangle -\beta_0^2}{\beta_0^2}= \frac{2}{\mu (n)}.
    \label{T-fluc}
    \end{equation}
This formula yields at the measured experimental density $n = 4.8 \cdot 10^{14}cm^{-1}$ {(the red data point on the left in Fig.~2)}  
the value $\mu \approx 10.5$ and thus $\delta \beta/ \beta= \delta T/ T \approx 0.44$. The authors of \cite{Wildnature} do not provide
precise temperature or temperature fluctuation measurements but
just mention in their paper that the temperature of the ions in their experiments is roughly given
given by
$T=(15\pm 5)$ Kelvin, i.e. they indicate a rather large uncertainty in temperature given by $\pm 5$ Kelvin. Apparently the order of magnitude of their stated $\delta T/T \approx 5/15=0.33$ coincides with the order of magnitude predicted by our superstatistical theory. In fact, ion trap experiments are usually plagued by large uncertainties in temperature, which we have now identified in this work as a fundamental
nonequilibrium quantum effect, rather than an imperfection of the experiment. 
In a sense, our theory presented is a {classical} theory of temperature fluctuations, which is relevant for high-density {quantum} reactant systems of small volume.
The predictions of our theory can be checked by ion-trap experiments where the reactants exhibit quantum-tunneling reactions of very small reaction rates,
so that very high densities are necessary to trigger the quantum-tunneling process.

In general, 
each realistically achievable ion trap experiment will have its own
equation of state $q=q(n)$, which could be experimentally measured.
A more general theory, derived in the End Matter section, predicts the formula
\begin{equation}
    q-1=\frac{2}{\mu (n) +4} \label{q-general-a}
\end{equation}  
with
\begin{equation}
    \mu (n)= C_1 \cdot \left( \frac{n_{max}}{n} \right)^\frac{2}{3} +C_0 \label{xxx}
\end{equation}
where $C_1$ and $C_0$ are two non-universal constants that depend on the experimental set-up.
Also the temperature fluctuation statistics could and should be investigated in future experiments.
The experimental details enter into the value represented by
the constants $C_0$ and $C_1$, which are non-universal and which depend e.g.
on the Mathieu stability parameters of the Mathieu differential equation describing the ion trap, as well as the mass ratios of the neutral particles and ions.
However, the exponent $2/3$ describing the area law is expected to be universal.
The function $q(n)$ with the given $C_0$ and $C_1$, in turn, fixes the statistics of the temperature fluctuations in the experimental system
via eq.~(\ref{T-fluc}). The effective entropy $S_q$ describing this scale-dependent
system has non-additive properties (see End Matter section for more details).

To conclude, in this Letter we have developed a generalized statistical mechanics theory
based on a superstatistical approach that gives quantitative predictions
how the entropic index $q$ depends on the density $n$ of reactants in slow quantum-tunneling reactions. Future ion trap experiments, varying in a systematic way the very high density of reactants, will be able to quantitatively test the predictions of our theory with high precision. The index $q=q(n)$ is relevant for both the shape of the anomalous velocity distributions as well as the statistics of temperature fluctuations in the system. 
\section*{End Matter}

\subsection*{Type-A versus Type-B superstatistics}

The averaging of the Boltzmann factors can be done in two different ways, either excluding the normalization constant $Z(\beta)$ or including it. This is usually referred to as type-A or type-B superstatistics \cite{BeckCohen2003}. 
For type-A superstatistics one simply considers averaged (unnormalized) Boltzmann factors
of the form
\begin{equation}
    B(E) =\int_0^\infty f(\beta) e^{-\beta E}d\beta .
\end{equation}
In our paper here, however, we use type-B superstatistics which is different from type-A superstatistics. Type-B is generally more physical since it basically corresponds to an appplication of Bayesian statistical methods for the local statistical mechanics, taking into account the proper normalization. Given some probability density $f(\beta)$ of inverse temperatures $\beta$ and some energy level $E$ corresponding to an arbitrary Hamiltonian $H$ one defines marginal probability densities in type-B superstatistics as
\begin{equation}
    p(E)=\int_0^\infty f(\beta) \frac{1}{Z(\beta)}e^{-\beta E}d\beta \label{here-it-is}
\end{equation}
where $Z(\beta)$is the partition function of the local statistical mechanics.
For $f(\beta)$ all kinds of probability densities can be used, but $q$-statistics as observed in ion trap experiments requires the $\chi^2$-distribution
\begin{equation}
    f(\beta) = \frac{1}{\Gamma(\frac{\mu}{2})} \left( \frac{\mu}{2 \beta_0} \right)^{\frac{\mu}{2}} \beta^{\frac{\mu}{2}-1} \exp
    \left( - \frac{\mu}{2} \frac{\beta}{\beta_0} \right) \label{fb}
\end{equation}
where
\begin{equation}
    \beta_0= \langle \beta \rangle = \int_0^\infty \beta f(\beta) d\beta =\mu \langle X_i^2 \rangle
\end{equation}
is the average inverse temperature in the system.
One immediately recognizes that the integration result in eq.~(\ref{here-it-is}) is influenced both by the functional form of $f(\beta)$ as well as what the local partition function $Z(\beta)$ of the local statistical mechanics is.
If we assume that the $\beta$-dependence
of $Z(\beta)$ is given by the functional form
\begin{equation}
    Z(\beta) \sim \beta^x e^{-y\beta}
\end{equation}
then, by explicit integration in eq.~(\ref{here-it-is}) using $f(\beta)$ as given in eq.~(\ref{fb}),
one obtains
\begin{equation}
    p_q(E)=\int_0^\infty d\beta f(\beta) \frac{1}{Z(\beta)} e^{-\beta E} \sim  e_q^{- \tilde{\beta} E}, \label{q-Boltzmann}
\end{equation}
where
\begin{equation}
    q=1+\frac{2}{\mu -2x} \label{qq}
\end{equation}
and
\begin{equation}
    \tilde{\beta}= \frac{\beta_0}{1+(q-1)(x-\beta_0y)}. \label{bqq}
\end{equation}
For the ion trap dynamics, local Maxwellian statistics $Ee^{-\beta E}/Z(\beta)$ with different temperatures was assumed, which led to the normalization  constant $Z(\beta)=\beta^{-2}$. Hence $x=-2$ and $y=0$ for the formulas
used in the main part of the paper.

\subsection*{Generalizing the scaling theory of $q(n)$}

Our generalized statistical mechanics theory developed in the main section opens the door to various intriguing new ideas and predictions that can be experimentally tested with quantum tunneling experiments of the type performed by Wild et al. \cite{Wildnature}. 
The first observation is that our predicted $n$-dependence in eq.~(\ref{qn}) is universal,
i.e. it does only depend on the density $n$, and not on further parameters.
This universality is expected to hold for an idealized experiment under
optimum conditions. Realistically, there will be many influences
that disturb this idealized setting. To take these effects into account,
we may combine our universal formula with the low-density kinetic theory
of ion traps developed by Rouse and Willitsch \cite{Rouse2017},
by allowing for a non-universal
additive constant in eq.~(\ref{mu}). In other words, we may generally write for the degrees of freedom
\begin{equation}
    \mu (n) = 3 \cdot \left( \frac{n_{max}}{n} \right)^\frac{2}{3} +C_0 \label{31}
\end{equation}
where the additive constant $C_0$ is non-universal and depends on the particular properties
of the experiment, such as the mass ratio $m_n/m_i$ or the Mathieu stability parameters of the ion trap. By this, the fundamental area-dependence law of the degrees of freedom,
reflected by the exponent $\frac{2}{3}$, is still conserved. This area dependence
somewhat reminds us of the area dependence of Bekenstein-Hawking entropy relevant for black holes, although this is only a formal analogy. In our case, the area dependence comes from the fact that heat is lost through the surface of the volume that contains the reactants.
Note that area laws (i.e. exponents 2/3) are also relevant in the general context of entanglement entropies \cite{eisert2010, CarusoTsallis2008}. The area dependence of the heat loss through the surface proposed in our approach here is thus consistent with the
area dependence of generalized entropies describing quantum entanglement, and in general it also allows for the embedding of more general entropies describing quantum processes \cite{jizba2019, regula2024}.

Once we allow the additive constant $C_0$ to take on non-universal
values that depend on the experimental details, there is no reason to keep the multiplicative constant 3 in front of the density-dependent term
in eq.~(\ref{31}). In fact, every experiment performed will have a different maximum density $n_{max}$ that can be achieved. In a sense the chosen parameter $n_{max}$ just sets a scale to which other densities are
compared to. Thus,
if we fix somewhat arbitrarily some density scale $n_{max}$, for different experiments the multiplicative constant in front of $(n_{max}/n)^{2/3}$
will be different and will depend on the experimental details. Hence most generally we may allow for a relation of the form
\begin{equation}
    \mu (n)= C_1 \cdot \left( \frac{n_{max}}{n} \right)^\frac{2}{3} +C_0 \label{32}
\end{equation}
where $C_1$ and $C_0$ are non-universal constants depending on the experimental set-up. 
We still expect the approximate relation $C_0+C_1 \approx 4$ to hold, since at the
largest experimentally achievable densities $n_{max}$ there are 4 space-time degrees of freedom available for the heat losses.
Formula (\ref{32}) still conserves the area-law of the degrees of freedom, described by the exponent 2/3. The above function $\mu(n)$ then fixes, as before, the parameter $q-1$ as
  \begin{equation}
    q-1=\frac{2}{\mu (n) +4} \label{q-general}
\end{equation}  
and this $\mu (n)$ also enters into the temperature fluctuation statistics as given by eq.~(\ref{T-fluc}).
We obtain an optimum fit of the data of the experiment of Wild et al. \cite{Wildnature,private} if we choose $C_1=2.14$ and $C_0=2.56$. This curve is shown in red in Fig.~2. 
Given some experimentally extracted $C_0$, there is then a more stringent bound on the maximum value of $q$ that can be experimentally achieved under the given non-universal experimental conditions,
namely $q_{max}=1+\frac{2}{C_0+4}$. 

\subsection*{Non-additivity of $S_q$ in the given experimental context}
The present approach also allows us to give physical meaning to the nonadditivity relation of the entropic functional $S_q$ 
\begin{equation}
S_q(A+B)=S_q(A)+S_q(B)-(q-1)S_q(A)S_q(B) \label{nonad1}
\end{equation}
which is generally satisfied for statistically independent subsystems $A$ and $B$ (see Supplementary Material for more details). Here the $q$-entropy $S_q$ is defined 
for any $q$ as \cite{Tsallis1988,Tsallis2023book,beck-review}
\begin{equation}
S_q=\frac{1}{1-q}(1 -\sum_ip_i^q)
\end{equation}
and it reduces to the BG-entropy $S_1=-\sum_i  p_i \ln p_i$ for $q \to 1$. The $p_i$ are the probabilities of the microstates $i$.

As the velocity distributions observed in the ion trap experiment effectively maximize the entropic functional $S_q$
subject to suitable constraints, one may ask what the physical meaning of the relation (\ref{nonad1}) is in the given experimental context. As evidenced by
the observed anomalous velocity distributions, we have $q>1$ and hence the term proportional to $q-1$ cannot be neglected.

Our physical interpretation is as follows. We note that our scaling theory implies a temperature fluctuation statistics
that is different in volumes $V$ of different size: Volumes of bigger size have smaller temperature fluctuations. If we put together
two subsystems $A$ and $B$ we get a different temperature fluctuation statistics (less fluctuations)
as the joint system $A+B$ has bigger size. Hence, for the effective entropy describing the system there must be a negative correction term for the joint system as compared to the sum of the entropies of the two smaller subsystems. This is the physical meaning of the relation  (\ref{nonad1}).

Apparently $S_q$ is non-additive \cite{Penrose1970} for any given $q$ different from 1, but if the given coupling term of strength $(q-1)$ is introduced, than the corresponding interacting field theory for the entropy is well-defined at {\em any} scale, and effectively takes into account
temperature and quantum fluctuations that take place in a volume-dependent way. For $q$-statistics of the type discussed in this paper we have a bilinear
interaction term
of strength $-(q-1)S_q(A)S_q(B)$ (similar to a dot-product of entropic vectors for the smaller subsystems) when combining two independent
subsystems $A$ and $B$ into a bigger subsystem $A+B$ where the temperature
statistics is different due to the larger volume. The non-additivity relation of $S_q$ as given by eq.~(\ref{nonad1}) indeed represents one of the simplest interacting field theories
of entropy one can think of, where temperature fluctuations are described in an effective way, although other types of superstatistics \cite{BeckCohen2003,BCS} or other types of entropies \cite{jizba2019,beck-review,regula2024,eisert2010,CarusoTsallis2008} could be also relevant.
For example, Rouse and Willitsch provide evidence in their paper \cite{Rouse2017} that
at absolute temperature $T=0$ log-normal superstatistics is relevant in ion traps rather than
$\chi^2$ superstatistics. For small but finite temperatures $T$ of the buffer gas, however, $q$-statistics of the type discussed here is in best agreement with the quantum tunneling experiment performed at high densities. Boltzmann-Gibbs, i.e. the case $q=1$, then simply describes a non-interacting field theory for entropy, this case is recovered for density $n \to 0$.
The experimental validation of these and/or similar features, i.e. 
systematic measurements of the equation of state $q=q(n)$ and of the temperature fluctuation statistics would be very welcome in future experiments. 

{\subsection*{Universality of the results}
Our theory predicts the existence of fundamental universal temperature fluctuations in a nonequilibrium situation that can be experimentally realized and tested by the latest generation of ion trap experiments. These temperature fluctuations are a fundamental effect and need to be distinguished from imperfections of the measuring device. The predicted function $q(n)$ is experimentally measurable and thus the results of our theory are experimentally testable. Universality kicks in at high densities, where the 4 space-time dimensions essentially fix the 4 degrees of freedom that are relevant for the $q$-Gaussian velocity distributions at the scale $n_{max}$. But also at low densities, where quantum tunneling is the dominant effect to enable the chemical reaction, our theory is applicable, as it predicts that in this case the parameter $q-1$ scales as $n^{2/3}$ if the density $n$ goes to zero.} 

\newpage 
\subsection*{Acknowledgements}
The authors thank R. Wester and R. Wild for kindly providing information regarding their paper and for fruitful discussions concerning the error bars of the index $q$. We also thank
H.S. Lima for useful discussions. C.B. acknowledges
funding from a QMUL ISPF Institutional Support Grant. Partial support from CNPq and Faperj (Brazilian agencies) is acknowledged as well.

\end{document}